\documentstyle[times,prl,aps,multicol,epsfig,graphics]{revtex}
\def\d{{\rm d}}
\def\tv{{\bf t}}
\def\bv{{\bf b}}

\begin{document}
\title{Statistical mechanics of double-helical polymers}
\author{Alvise De Col$^{1,2}$ and Tanniemola~B. Liverpool$^{1}$}
\address{${}^1$ Condensed Matter Theory Group, Blackett Laboratory, Imperial College,
London SW7  2BZ\\
        ${}^2$ Theoreticshe Physik, ETH H\"onggerberg, CH-8093 Z\"urich, Switzerland } 
\date{\today}
\maketitle
\begin{abstract}
  We introduce a simple geometric model for a double-stranded and
  double-helical polymer. We study the statistical mechanics of such
  polymers using both analytical techniques and simulation. Our model
  has a single energy-scale which determines both the bending and
  twisting rigidity of the polymer. The helix melts at a particular
  temperature $T_c$ below which the chain has a helical structure and
  above which this structure is disordered. Under extension we find that
  for small forces, the behaviour is very similar to worm-like chain
  behaviour but becomes very different at higher forces.

\end{abstract}
\begin{multicols}{2}
  Recent developments in single molecule manipulation techniques has
  led to the detailed study of the mechanical properties of DNA and
  other biomolecules as well as their response to applied fields.  The
  model most used in the study of the large-scale properties of
  biopolymers is that of the worm-like chain~\cite{KratPor} in which
  the polymer flexibility (structure) is determined by a single
  length, the persistence length $L_p$ which measures the
  tangent--tangent correlations. For example, DNA has a persistence
  length $L_p \approx 50\mbox{nm}$.  Such coarse-grained models are
  needed to understand the {\em statistical} behaviour of long chains
  with a {\em large} number of monomers. They are complementary to
  chemically specific models which describe accurately the {\em local}
  effects of external fields but generally cannot deal with long
  chains~\cite{chem_dna}.  Generalisations of the worm-like chain to
  introduce twist degrees of freedom have also been
  proposed~\cite{Marko}. Whilst the worm-like chain model and its
  generalisations give a good account of DNA under small fields
  (perturbations), it fails when these perturbations become large,
  e.g. under tensional forces above $65$pN~\cite{Smith,Cluzel}.  Such
  situations are biologically relevant in for example in DNA
  replication and repair.

  DNA is double-stranded and helical. We introduce a geometric model
  which includes three important ingredients of DNA in B-form.  First,
  the double-stranded nature of DNA given by the two phosphate
  backbones; second, the hydrogen bonds that keep the two strands
  together and third, the interactions between the bases that bring
  about the twisted structure.  Similar models could be used to study
  other double-helical polymers like F-actin.  The model is a
  generalisation of the double-stranded semiflexible (ribbon) polymer
  introduced by Liverpool, Golestanian and Kremer (LGK)~\cite{LGK}
  which takes into account the first two aspects but ignores the
  third. Other ribbon models have also been proposed by a number of
  authors~\cite{ribbon}. It will be our conclusion that it is exactly
  this third property, the base-stacking interaction, which can
  account for many of the {\em elastic} properties of DNA.
Unlike Z.  Haijun et al~\cite{zhou}, who proposed another
generalisation of the LGK model with base-stacking interactions, we
find that the {\em rigidity} of the individual strands is irrelevant
for the effective persistence length, $L_p$.
This is consistent with the large difference between the persistence
length of single and double-stranded DNA measured experimentally.
Our model can easily be extended to include excluded volume and
electrostatic interactions which for the moment have been
ignored~\cite{project}.
We do not attempt to make quantitative comparison with experiments but
point out qualitative differences between simpler models.  Our aim is
rather to suggest a minimal model required to understand experimental
results.

\begin{figure}
\begin{center}
\epsfig{file=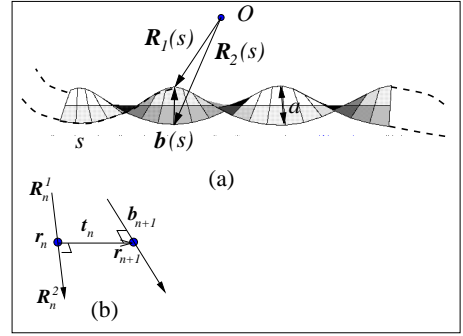,width=6cm}
\end{center}
\caption{\protect (a) The schematic double-helical polymer with ${\bf R}_2(s)={\bf R}_1(s) + a {\bf b}(s)$. The mid section (solid) is in the minimum energy configuration. (b) The discretisation of the double-helix used in the simulation. }
\label{fig1}
\end{figure}

  The polymer is embedded in three dimensional space. As for the
  ribbon, we see qualitatively different behaviour for the high and
  low temperature regimes with a {\em melting} of the helix at a
  particular temperature $T_c \simeq \left({4 \over 9}\right)^2
  \left[{ 5 + (2 P / \pi a)^2 \over 1 + (2 P / \pi a)^2}
  \right]^{3/2}B /k_B a $ where $P$ and $a$ are respectively the pitch
  and diameter of the helix at $T=0$ and $B$ the `stiffness' of the
  base-stacking interaction. Below $T_c$, there is a helical structure
  of the chain whilst above $T_c$, the helical structure vanishes.
  Unlike the ribbon, {\em both} the bend and twist rigidity of the
  double-helical polymer are determined by the base-stacking
  interaction and the stiffness of the individual strands is
  irrelevant. We also study the response of the
  double-helical polymer to an extensional force and find that for
  small forces, the mechanical response is very similar to that of a
  worm-like chain but becomes significantly different at higher
  forces.

The system is composed of two semi-flexible chains modelling the sugar
phosphate backbones, each with rigidity $A$, whose embeddings in
$3$-dimensional space are defined by ${\bf R}_1 (s)$ and $ {\bf R}_2
(s)$.
Because of the covalent bonds ($E_{Cb} \sim 100 k_B T$ at room
temperature) between the alternating phosphate and sugar molecules,
the backbones are effectively inextensible at the forces and
temperatures we consider.
Similarly, the H-bonds
($E_{Hb} \sim 10 k_B T$)
between the two backbones are assumed to keep the
distance between the chains constant.

The separation of the strands is fixed and given by $a$, and the
H-bonds give the constraint ${\bf R}_2 (s)={\bf R}_1(s)+ a {\bf b}(s)$
where ${\bf b}(s)$ is a unit vector which we call the {\em
bond-orientation vector} (see Fig.~\ref{fig1}).  The bond-orientation
vector is perpendicular to both strands.  The infinitesimal distance
in $3$-dimensional space between neighbouring points on a single
strand is a function of the {\em local} curvature , $\partial_s^2 {\bf
R}_{i}(s)$ of the strands, where $\partial_x A(x)\equiv \partial A /
\partial x$. The base-stacking interaction can be modelled
by a potential $V_{i}(\partial_s^2 {\bf R}_{i},{\bf b})$, where the
subscript $i \in \{1,2\}$ refers to strand 1 or 2, which is a function
of the curvatures of the chains and the bond-vector only, with a
minimum given by a symmetric double-helix.  The Hamiltonian of the
system is given by

\begin{eqnarray}
{\cal H} = \int ds \left\{
    V_1(\partial_s^2 {\bf R}_{1},{\bf b}) + V_2(\partial_s^2 {\bf
    R}_{2},{\bf b})\right. \nonumber\\
    \left.+ A(|\partial_s^2 {\bf R}_1|^2+ |\partial_s^2 {\bf R}_2|^2) \right\} \, .
\end{eqnarray} 

The simplest example of such
    a potential which we will use for the remainder of this letter is
    the quadratic $V_1(\partial_s^2 {\bf R}_{1},{\bf b})=
    B(\partial_s^2 {\bf R}_{1}-H{\bf b})^2$ and $V_2(\partial_s^2 {\bf
    R}_{2},{\bf b})= B(\partial_s^2 {\bf R}_{2}+ H{\bf b})^2$.  This
    is a potential of {\em stiffness} $B$ with a minimum when
    equivalent points on the two strands have equal and opposite
    curvatures, $\pm H$.  More complicated potentials with several
    minima are also possible (see later).  We note that we have not
    included the asymmetry of the DNA helix but a similar but more
    complicated model could be defined for an asymmetric
    double-helix~\cite{project}.  We implement the model by
    introducing the ``mid-curve'' ${\bf r}(s)$: $ {\bf R}_1 (s) = {\bf
    r}(s)+\frac{a}{2} \; {\bf b}, \, \, \, {\bf R}_2 (s) = {\bf
    r}(s)-\frac{a}{2} \; {\bf b}.  $ In terms of the tangent to the
    mid-curve ${\bf t}(s)=\partial_s {\bf r}$ and the bond-director
    ${\bf b}$ the Hamiltonian of the system can now be written as

\begin{eqnarray}
{\cal H}[{\bf t},{\bf b}]=\frac{B}{2} \; \int \d s \; \left[
\left(\partial_s {\bf t} (s) + {a\over 2} \partial_s^2 {\bf b} (s)- H {\bf b}\right)^2\right. \nonumber \\ 
+\left.\left(\partial_s {\bf t} (s) - {a\over 2} \partial_s^2 {\bf b} (s) + H {\bf b}\right)^2
\right],\label{Htn}
\end{eqnarray}

subject to the exact (local) constraints
\begin{eqnarray}
({\bf t} \pm \frac{a}{2} \;\partial_s {\bf b})^2=1, \qquad
{\bf b}^2=1, \qquad
({\bf t} \pm \frac{a}{2} \;\partial_s {\bf b}) \cdot {\bf b}=0.
\label{constr}
\end{eqnarray}
corresponding to the chain inextensibility, fixed distance between
strands and definition of the bond vector respectively.  We make the
assumption that $B \gg A$ since the bending rigidity of
single-stranded DNA is very small, ($A/k_B T < 10\AA$), compared to
the double-stranded form.  We can define a typical length $\ell =
B/k_B T$, which together with the strand separation $a$ and
equilibrium radius of curvature $H^{-1/2}$ form the three relevant
lengths of the problem. This completes the formulation of the model.

\paragraph*{T=0 behaviour}
We can calculate the ground state of this system from the condition
${\cal H}=0$ which gives, ${\bf R}_1(s)=\frac{a}{2}\cos\left(\Theta
  s\right)\hat{e}_1 + \frac{a}{2}\sin\left(\Theta\,s\right)\hat{e}_2
+s\sqrt{1-\frac{a^2\Theta^2}{4}}\,\hat{e}_3 $ and ${\bf
  R}_2(s)=-\frac{a}{2}\cos\left(\Theta\,s\right)\hat{e}_1-
\frac{a}{2}\sin\left(\Theta\,s\right)\hat{e}_2
+s\sqrt{1-\frac{a^2\Theta^2}{4}}\,\hat{e}_3 $ where $\Theta=
\sqrt{\frac{2H}{a}}$. This is a double helix with a pitch: $P={2\pi
  \over \Theta}\sqrt{1-\frac{a^2\Theta^2}{4}}$ and the bond-director
field, $ {\bf b}(s)=-\cos\left(\Theta\,s\right)\hat{e}_1-
\sin\left(\Theta\,s\right)\hat{e}_2$ where $\{\hat{e}_i\}$ are an
orthonormal set of unit vectors.

\paragraph*{Finite T behaviour}
Inspite of the simplification of the model from using constraints, an
exact analytic expression for the partition function of the model is
still not available. For our analytic calculations, we use a very
useful approximation~\cite{semiflex} which describes correctly the
equilibrium behaviour of correlation functions and even distribution
functions, if the end effects are taken into account
carefully~\cite{wlc_mean}. This is a mean field approach that relaxes
the local constraints to global ones\cite{semiflex,wlc_mean}. It may
be thought of as a self-consistent theory and corresponds to the
saddle-point evaluation of path integrals over the lagrange
multipliers\cite{David}. This approach fails for some dynamical
properties~\cite{morse} which are outside the scope of this letter.
To implement this, we add a variational term to our Hamiltonian $ \frac{{\cal
H}_{m}}{k_{B}T} = \int \d s \left[ \frac{b}{\ell}({\bf t}-\frac{a}{2}
\;\partial_s {\bf b})^2 +\frac{b}{\ell}({\bf t}+\frac{a}{2}
\;\partial_s {\bf b})^2 +\frac{c a^2}{4 \ell^3} {\bf b}^2 \right .$
$\left . + \frac{e}{\ell} ({\bf t}-\frac{a}{2} \;\partial_s {\bf b})
\cdot {\bf b} +\frac{e}{\ell} ({\bf t}+\frac{a}{2} \;\partial_s {\bf
b}) \cdot {\bf b} \right],\nonumber $ where $b$, $c$, and $e$ are
dimensionless constant Lagrange multipliers .

Two useful geometric parameters $u=H a/2$ and $v=4 (\ell/a)^2$ can be
used to characterise the double-helical polymer.  Note that $v$ is
proportional to $T^{-2}$ and can be viewed as a measure of
temperature.  We then determine the constants self consistently by
demanding the constraints of Eq.(\ref{constr}) to hold on average,
where the thermal average is calculated by using the total Hamiltonian
${\cal H}+{\cal H}_m$.  The self-consistency lead to the following set
of equations for the constants $b$ and $c$: $\left(c+u^2
  v^2\right)\left[(b-u v)+\sqrt{c+u^2 v^2}\right]=\frac{9 \,v^2}{32} ;
\frac{1}{4\sqrt{2b}}+\frac{1}{3}\frac{1}{v}\sqrt{c+u^2
  v^2}=\frac{1}{3}$ and $e=0$. The above equations, which are
nonlinear and difficult to solve exactly, determine the behaviour of
$b$ and $c$ functions of $u$ and $v$.  One can solve the equations
analytically in some limiting cases corresponding to $T \rightarrow 0$
and $T \rightarrow \infty$.  For $v\rightarrow\infty$
($T\rightarrow0$) and $H\neq0$ we find,
\begin{math}
b\sim\frac{9}{32}\frac{1}{(1-u)^2} ,
c\sim\frac{9}{16}\left[\frac{1}{u^2}-\frac{1}{(1-u)^2}\right]u v
\end{math} and for
$v\rightarrow0$ ($T\rightarrow\infty$) and $H\neq0$, we obtain
\begin{math}
b\sim\frac{9}{8} ,
c\sim\left(\frac{1}{4u^2}-1\right)u^2 v^2
\end{math}
A full solution requires a numerical treatment
and shows simple monotonic behaviour for both $c$ and $b$~\cite{inprep}.

We can then calculate the correlation functions.  For the
tangent--tangent correlation one obtains
\begin{eqnarray}
\left<\tv(s)\cdot\tv(0)\right>
                    & = & \frac{3}{4\sqrt{2b}}e^{-\sqrt{2b}\frac{s}{\ell}}
                    \stackrel{T\rightarrow0}{\approx}  (1-u)e^{-\frac{3}{4(1-u)}\frac{s}{\ell}}
\label{corrt0}
\end{eqnarray}
whereas for the bond-director field one obtains
\end{multicols}
\begin{eqnarray}
\langle\bv(s)\cdot\bv(0)\rangle
                    & = & \frac{3i}{2a^2}\frac{\ell^2}{\ell^2\sqrt{D}}
    \left[\frac{e^{i\sqrt{(u v-b)+\sqrt{D}\ell^2}\frac{s}{\ell}}}{\sqrt{(u v-b)+\sqrt{D}\ell^2}}+\frac{e^{-i\sqrt{(u v-b)-\sqrt{D}\ell^2}\frac{s}{\ell}}}
              {\sqrt{(u v-b)-\sqrt{D}\ell^2}}\right]
      \stackrel{T\rightarrow0}{\approx} e^{-\frac{3}{8u}\frac{s}{\ell}}\cos\left(
        \sqrt{\frac{2H}{a}}s\right)
\label{corrb0}
\end{eqnarray}
\begin{multicols}{2}
  We have defined a discriminant $D(u,v)=\frac{1}{\ell^2}\sqrt{(b-u
    v)^2-(c+u^2 v^2)}$. The tangent--tangent correlation
  (Eq.(\ref{corrt0})) of the double-helical polymers is similar to
  that of a worm-like chain but with a length rescaling factor, and
  decays exponentially at all temperatures.  Eq.(\ref{corrb0}) on the
  other hand, indicates a change of behaviour at $D=0$ i.e.  $(b- u
  v)^2=c+u^2 v^2$ for the bond-director correlation.  The correlation
  is {\it over-damped} for $(b- u v)^2 > c+u^2 v^2$ (high
  temperatures), while it is {\it under-damped} (oscillatory) for
  $(b-u v)^2 < c+u^2 v^2$ (low temperatures). The interesting point
  $(b-u v)^2=c+u^2 v^2$ happens for $v_c \simeq 9 \left({9 \over 4 (1
      + 4u)}\right)^3$, that leads to the value for $T_c$ quoted
  above.  We also find a divergence in the specific heat, $C_{V}$ at
  $T_{c}$.  We emphasise that it is not a thermodynamic phase
  transition in the sense of long-range ordering and broken symmetry.
  It is a cross-over that appears due to competing effects, and the
  transition is from a state with some short-range order to a state
  with a different short-range order.  The cross-over ({\it
    transition}) point corresponds to a type of {\it`Lifshitz point'}
  for a 1-d system~\cite{Hornreich}.  At high temperatures, in
  addition to the effects we describe we also expect the H-bonds
  between the bases to break, leading to {\em denaturation} which we
  cannot treat in this model. In the low $T$ regime, we can read off
  the bend (tangent) persistence length $L_{TP} = {4 \over 3} \ell ( 1
  - H a/2)^{-1} $ and twist (bond) persistence length, $L_{BP} = {4
    \over 3} \ell H a $ and helical frequency $(2 H /a)^{1/2}$. Both
  the bend and twist rigidity of the double-helical polymer are given
  by the same energy scale $B=k_BT \ell$ and have little to do with
  the stiffness of the individual strands.

\paragraph*{Force-extension: unwinding the helix}
Under a stretching force, which without loss of generality we orient
along the z-axis ${\bf F}=F\hat{\bf z}$, the Hamiltonian becomes
${\cal H}_F = {\cal H} - {\bf F}\cdot \int ds \partial_s {\bf r}$. Now
$\langle t_z \rangle = \langle {\bf t} \cdot \hat{\bf z} \rangle \neq
0$ and we must use connected correlation functions to calculate the self-consistent equations. Defining $f =
Fa/8$ we obtain the equations
\begin{equation}
\left\{\begin{array}{l}
\frac{1}{4\sqrt{2b}}+\frac{1}{3}\left(\frac{f}{b}\right)^2{v}+\frac{1}{3}\frac{1}{v}\sqrt{c+u^2 v^2}=\frac{1}{3}\\
  \left(c+u^2 v^2\right)\left[(b-u v)+\sqrt{c+u^2 v^2}\right]=\frac{9 v^2}{32}
 \end{array}
 \right.
\label{fdif0}
\end{equation}
We can solve eqns. (\ref{fdif0}) for $b(u,v,f)$ and $c(u,v,f)$ and the
extension per unit length $z(u,v,f)$ is simply:
\begin{equation}
z=\frac{1}{L}\int_{0}^{L}\langle t(s)\rangle\,ds = \frac{f}{b}\sqrt{v}
\end{equation}

\begin{figure}
\begin{center}
\epsfig{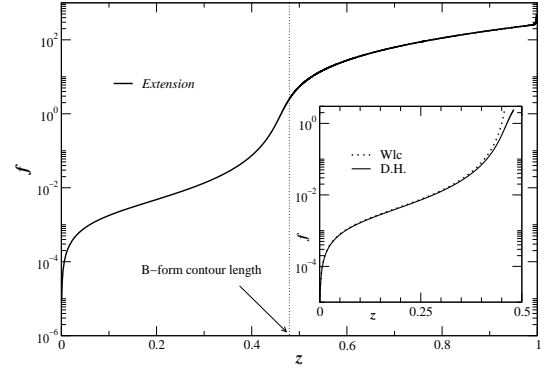}
\end{center}
\caption{\protect
  Force-extension relation for the double-helical polymer (DHP). This
  is a log plot with $u=0.77$ and $v=115000$, typical values for B-DNA
  which corresponds to a low temperature $T < T_c$ (Note
  $f=Fa/8$).The dashed line shows the $T=0$ contour length $\sqrt{1-a^2\Theta^2/4}$.
   Inset: comparison with worm-like chain at small forces.}
  \label{fig:ext}
\end{figure}

We can therefore obtain the force-extension relation for the
double-helical polymer. Eq.~\ref{fdif0} has been solved numerically
and in Fig.~\ref{fig:ext} we have plotted $z$ against $f$. In the inset we compare the force-extension curve to that of the WLC and find they are identical for small forces. We note the interesting fact that the mean
field model can also be used to calculate the force-extension
behaviour of the WLC.  It has the simple compact form $f L_p = {3 z
  \over 2 (1-z^2)^2}$. We find that this simple expression has an
error of less than 4 \% compared to the `exact' value of Marko and
Siggia~\cite{wlc_mean}.

\paragraph*{Simulations}

We also performed Monte Carlo simulations of the double-helical
polymers.  We used a square lattice of size $\Delta$ to discretise the
`helical ribbon' and to implement all the constraints (see
Fig.~\ref{fig1}). The discretised Hamiltonian is given by 
\begin{eqnarray}
\beta{\cal
  H} = {\ell} \sum_{n=1}^{N-1}\left({2 \over \Delta}+H^2\Delta-
  \sum_{j=1}^{2}\left[{\tv_{n+1}^j\cdot\tv_n^j\over \Delta}\right.\right. \nonumber\\
  \left.\left. +(-1)^jH\bv_n\cdot(\tv_{n+1}^j-\tv_{n}^j)\right]\right) \, ,
\end{eqnarray}
 where $n$ is the
monomer label (position along the chain) and $j \in \{1,2\}$ the
strand label (see Fig.~\ref{fig1}).  Since the Hamiltonian is local,
we grew chains with a local algorithm.
In our simulation the number
of monomers $N$ was taken to be 1000 for each chain, $\Delta=1$ and
$a=3$.
Typical
conformations at high and low temperatures are shown in Fig.
\ref{fig:conform}.

\begin{figure}
\begin{center}
\epsfig{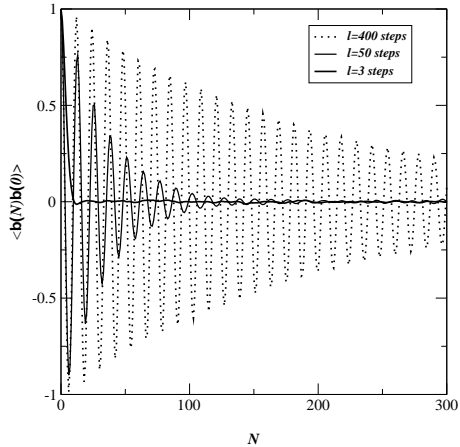}
\end{center}
\caption{The $\langle
  {\bf b}(s) \cdot {\bf b}(0)\rangle$ correlation function measured in
  the simulations for temperatures $\ell =400,50,3$ and $u=0.6$
corresponding to $T < T_c$ and $T>T_c$.
The averages were done over $200$ statistically independent samples.}
\label{fig:bb}
\end{figure}

\begin{figure}
\begin{center}
\epsfig{file=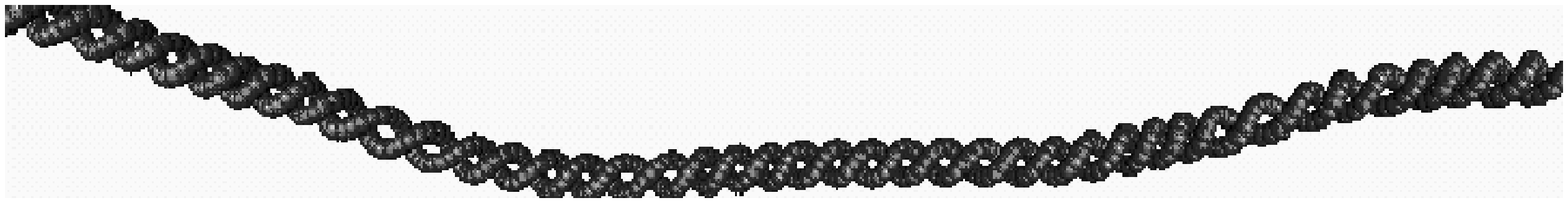,width=5.7cm}\newline
\epsfig{file=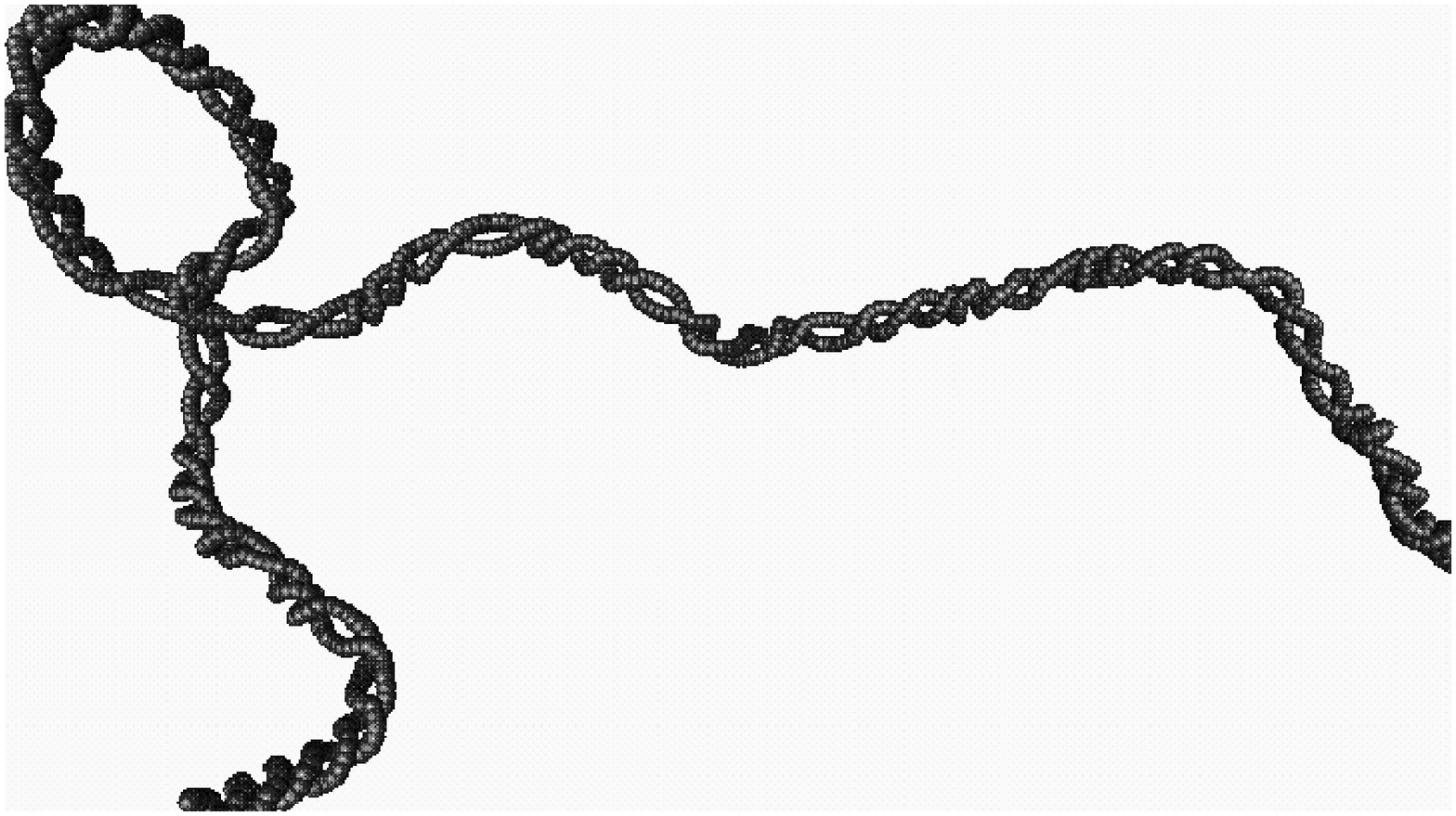,width=5.7cm}\newline
\epsfig{file=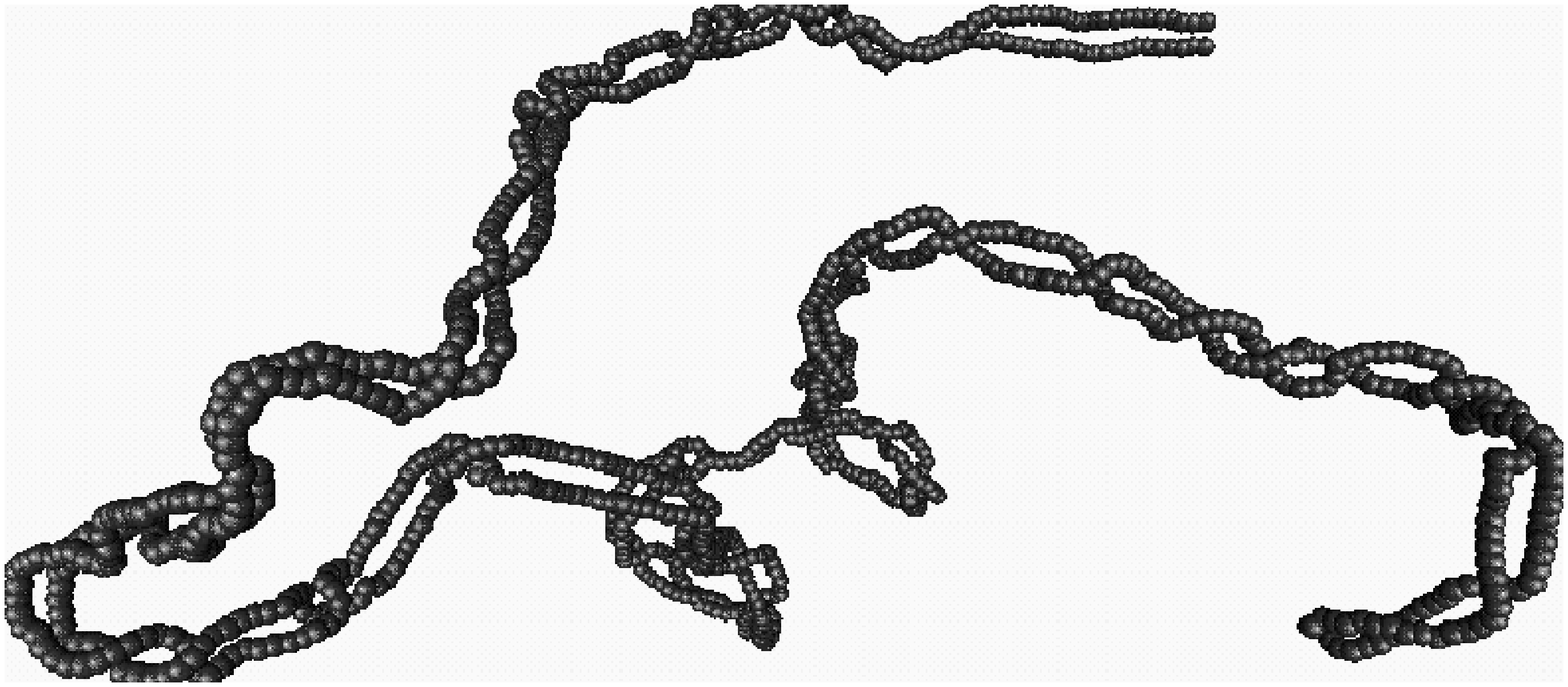,width=5.7cm}\newline
\end{center}
\caption{Typical conformations at (a) low, (b) intermediate 
and (b) high temperatures}
\label{fig:conform}
\end{figure}

Typical bond correlation functions are plotted in Fig.\ref{fig:bb}.
From the bond and tangent correlation functions, we can obtain the
bond and tangent persistence lengths. We calculate the bond
correlation length as $L_{BP}= C_B u \ell$ and the tangent correlation
length as $L_{TP}= C_T \ell$ with the constants given $C_B = 1.09 \pm
0.09$ and $C_T=2.46 \pm 0.73$ respectively. 
A comparison with the
mean field results shows that the results are the same to within a
constant. This difference can be considered as due to finite
corrections to the mean field solution due to higher order terms in
the $1/d$ expansion.

In conclusion, we have studied the properties of a well-defined model
of a double-helical, double-stranded semi-flexible polymer using a
mean-field analytical approach as well as extensive MC simulations.
We have shown novel non-trivial differences between the high, low {\it
  and} zero temperature behaviour.  A detailed comparison with the
simulations and the mean-field solution shows qualitative and almost
quantitative agreement.
The fine structure studied in this model gives a different behaviour
compared to a WLC.  We see this most clearly when the double-helical
polymer is subject to an external pulling force. It shows two distinct
regimes.  For low forces the dynamical response is simular to a WLC,
with the most relevant energetic contribution given by the bending
rigidity.  At higher forces the extension changes abruptly and
increases steeply as the double helical structure is {\em unwound}.
This \emph{qualitatively} agrees with the experimental results on
over-stretched DNA \cite{Smith,Cluzel}, which nevertheless have a
sharper transition and a flatter plateau.  The graph is plotted on a
log scale, so as to make clearer the complex behaviour of the
force-extension curve.  There are a two obvious ways in which a
sharper transition could be obtained.  First, we have used a simple
quadratic potential for the base-stacking interaction corresponding
only to an expansion about the minimum energy conformation.  A more
realistic potential with a cut-off and/or several minima could change
the sharpness of and number of transitions.  In addition one could
imagine as in mechanical unzipping experiments on DNA, that there
could be cooperative effects due to the effects of sequence disorder.

We have benefited from discussions with R.  Everaers, R. Golestanian
and K. Kremer.

\end{multicols}
\end{document}